\pdfoutput=1
\documentclass[10pt,twocolumn]{article}
\usepackage[utf8]{inputenc}
\usepackage[margin=0.68in,top=0.62in,bottom=0.65in]{geometry}
\usepackage{amsmath,amssymb,amsfonts}
\usepackage{booktabs}
\usepackage{graphicx}
\usepackage{hyperref}
\usepackage{cite}
\usepackage{microtype}
\usepackage{xcolor}

\makeatletter
\renewcommand\section{\@startsection {section}{1}{\z@}%
  {-1.8ex \@plus -0.5ex \@minus -.2ex}%
  {0.9ex \@plus.2ex}%
  {\normalfont\large\bfseries}}
\renewcommand\subsection{\@startsection{subsection}{2}{\z@}%
  {-1.4ex\@plus -0.4ex \@minus -.2ex}%
  {0.6ex \@plus .2ex}%
  {\normalfont\normalsize\bfseries}}
\renewcommand\paragraph{\@startsection{paragraph}{4}{\z@}%
  {1.0ex \@plus0.3ex \@minus.2ex}%
  {-0.8em}%
  {\normalfont\normalsize\bfseries}}
\makeatother

\hypersetup{
  colorlinks=true,
  linkcolor=black,
  citecolor=blue,
  urlcolor=blue
}

\title{\vspace{-2.0em}\textbf{\large Swanada: Split-Encoder Capability-Routed Conformer CTC and\\Sub-30~MB On-Device Speech Recognition Across Twelve Scripts}\vspace{-0.8em}}

\author{
  \textbf{Hemanth HM} \\
  \small\url{https://h3manth.com} \quad$\cdot$\quad \url{https://github.com/hemanth/swanada}
}

\date{\vspace{-1.5em}}

\begin{document}

\maketitle

\begin{abstract}
Client-side automatic speech recognition (ASR) across Indic and Classical scripts eliminates cloud transcription costs, network tail latency, and server-side audio retention, yet standard neural exports face four systems bottlenecks in web browsers and mobile runtimes: (1)~monolithic per-language Conformer checkpoints require $\sim$177~MiB per script ($> 2.1$~GiB across 12 scripts), exhausting browser \texttt{CacheStorage} quotas; (2)~exported relative positional encoding tables embed 10.2--20.5~MB of static FP32 constants while $1\times 1$ pointwise \texttt{Conv1d} operators trigger slow or unsupported \texttt{ConvInteger} kernels in WebAssembly SIMD; (3)~sliding-window CTC decoding flickers on boundary tokens as right context arrives; and (4)~mobile \texttt{WebAudio} pipelines suffer hardware sample-rate mismatches (44.1/48~kHz vs.\ 16~kHz) and silent \texttt{AudioWorklet} stalls under iOS \texttt{WKWebView} and Android backgrounding.

We present \textbf{Swanada}, an open-source 16~kHz streaming ASR system spanning Classical Sanskrit, Indian English, and ten Indian languages (\texttt{sa}, \texttt{en}, \texttt{hi}, \texttt{kn}, \texttt{mr}, \texttt{bn}, \texttt{ta}, \texttt{te}, \texttt{ml}, \texttt{gu}, \texttt{pa}, \texttt{or}) across browser WebAssembly, Node.js, and Python. Swanada pairs capability-routed shared INT8 Conformer encoders---a 17-layer 177.38~MiB Studio tier ($d = 512$, 115.2M parameters) for desktop/server runtimes and a distilled 14-layer 25.35~MiB Mobile tier ($d = 256$, 23.56M parameters) for phones---with twelve hot-swappable 515.0~KiB native-script linear CTC heads ($257 \times 512$), shrinking 12-language fleet storage by $11.6\times$ (183.4~MiB Studio) and $67.8\times$ (31.39~MiB Mobile; 25.85~MiB first load). A zero-dependency protobuf tail patcher exposes the pre-head representation in $< 2$~ms without shipping separate backbone files, while dynamic ONNX trigonometric positional synthesis, \texttt{MatMul}-backed pointwise convolutions, and post-quantization \texttt{QInt8} head calibration preserve sub-30~MB mobile accuracy. On $N = 480$ held-out studio utterances and $N = 1{,}280$ unseen public \texttt{test} clips across 16 open benchmarks (\texttt{FLEURS}, \texttt{LibriSpeech}, \texttt{GLOBE}, \texttt{AMI}, and wild Sanskrit), Swanada achieves 5.43\% held-out CER and 8.87\% 12-language open \texttt{FLEURS}+Wild CER on Studio (18.56\% and 19.23\% on 25.35~MiB Mobile) at $0.0097\times$--$0.0264\times$ 4-thread CPU RTF, delivering 30.6~ms median streaming TTFT with 0.00\% committed-prefix churn across 72 streaming sessions.
\end{abstract}

\section{Introduction}
\label{sec:intro}

Running speech-to-text directly inside a browser tab or local edge process removes per-minute API billing, round-trip network jitter, and transmission of private microphone audio to remote servers. Voice input for South Asian digital public infrastructure, Sanskrit recitation study, bilingual developer tools, and offline accessibility interfaces requires native script fidelity across Brahmic and Latin orthographies, sub-100~ms streaming partial updates, and reliable execution within smartphone browser memory budgets, complementing client-side speech synthesis~\cite{hemanth2026ninada}.

Convolution-augmented Transformers (Conformers)~\cite{gulati2020conformer} trained with Connectionist Temporal Classification (CTC)~\cite{graves2006ctc} are attractive for on-device ASR because frame-synchronous emission avoids the $O(U^2)$ autoregressive decoder loop and hallucination modes of sequence-to-sequence models such as Whisper~\cite{radford2023whisper}. However, deploying Conformer-CTC across twelve scripts inside real desktop and mobile browsers exposes four concrete systems bottlenecks:
\begin{enumerate}\setlength{\itemsep}{1pt}\setlength{\parskip}{0pt}
    \item \textbf{Monolithic Fleet Storage vs.\ Joint-Vocabulary Dilution}: Shipping one quantized Conformer-Large per language requires $12 \times 177.38 = 2{,}128.5$~MiB, exceeding mobile browser cache budgets. Conversely, pooling twelve scripts into a single flat output softmax ($> 3{,}000$ classes) inflates final projection compute by $12\times$ at every frame and permits cross-script token leakage during noisy frames.
    \item \textbf{ONNX Constant Bloat and \texttt{ConvInteger} Quantization Gaps}: Standard Conformer exports materialize a static $2 L_{\max} \times d$ FP32 relative positional encoding buffer (20.48~MB at $d = 512$, 10.24~MB at $d = 256$ for $L_{\max} = 5{,}000$), consuming over a third of a 30~MB mobile budget on uncompressible constants. Simultaneously, exporting $1\times 1$ pointwise convolutions inside Conformer macaron blocks as \texttt{Conv} nodes either leaves weights in FP32 or emits \texttt{ConvInteger} operators lacking fast WebAssembly SIMD kernels in \texttt{onnxruntime-web}.
    \item \textbf{Quantization Feature Drift Under Narrow Student Bottlenecks}: Compressing a 115.2M-parameter teacher into a $< 30$~MB student and applying dynamic INT8 weight/activation quantization perturbs top-layer hidden norms and directions, degrading pre-trained linear CTC decision boundaries unless explicitly recalibrated on quantized ONNX activations.
    \item \textbf{Streaming Prefix Flicker and Mobile \texttt{WebAudio} Fragility}: Evaluating bounded audio windows causes trailing subwords to oscillate as right acoustic context arrives, while mobile browsers lock \texttt{AudioContext} to hardware rates (44.1/48~kHz) and intermittently stall \texttt{AudioWorklet} threads under iOS \texttt{WKWebView} rules.
\end{enumerate}

\begin{figure*}[t]
\centering
\includegraphics[width=0.88\textwidth]{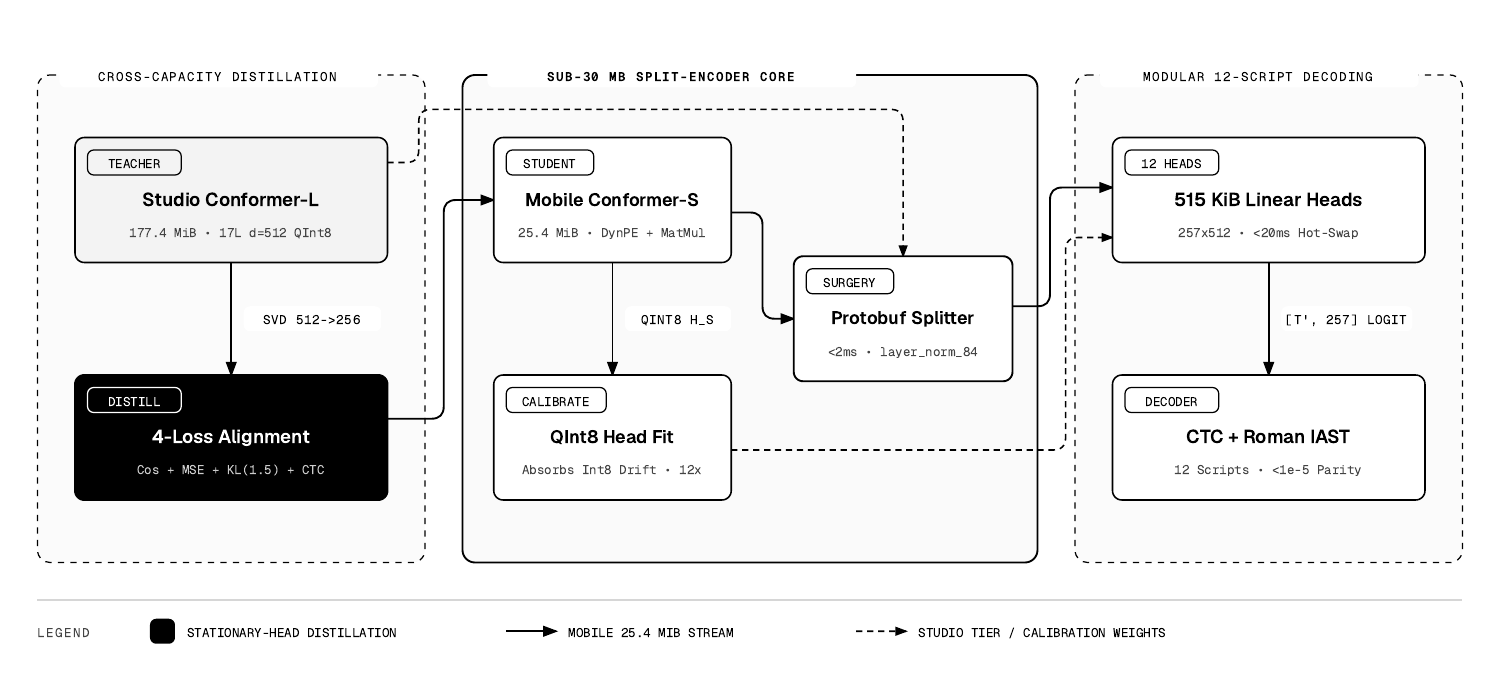}
\caption{Swanada split-encoder decomposition and cross-capacity distillation: the 17-layer Studio Conformer-Large ($d = 512$, 177.38~MiB \texttt{QInt8}) distills into a 14-layer Mobile Conformer-Small ($d = 256$, 25.35~MiB \texttt{QInt8}) with runtime trigonometric positional synthesis and \texttt{MatMul}-backed pointwise convolutions. Post-quantization head calibration fits twelve 515.0~KiB linear CTC heads to actual \texttt{DynamicQuantizeLinear} features, while a $< 2$~ms protobuf tail patcher exposes \texttt{layer\_norm\_84} at session load.}
\label{fig:distill}
\end{figure*}

To address these constraints across browser ES modules, Node.js, and Python without custom WebAssembly builds, we present \textbf{Swanada} (Figures~\ref{fig:distill} and~\ref{fig:pipeline}). Our contributions are threefold:
\begin{itemize}\setlength{\itemsep}{1pt}\setlength{\parskip}{0pt}
    \item \textbf{Runtime Protobuf Tail Surgery and Modular 12-Script CTC}: We decouple shared pan-Indic acoustic encoding from twelve script-isolated 257-token linear heads (515.0~KiB each), using a zero-dependency binary ONNX \texttt{GraphProto} patcher that exposes the shared $512$-dimensional encoder representation in $< 2$~ms and switches languages in $< 20$~ms with $< 10^{-5}$ logit deviation against monolithic graphs.
    \item \textbf{Sub-30~MB Mobile Conformer Distillation and \texttt{QInt8} Head Calibration}: We design a 25.35~MiB Mobile Conformer-Small combining zero-table dynamic relative positional encoding, \texttt{MatMul}-backed pointwise convolutions for 100\% SIMD \texttt{QInt8} coverage, stationary-hyperplane 4-loss distillation from the 177.38~MiB teacher, and post-export \texttt{QInt8} head calibration.
    \item \textbf{Two-Zone LCP Streaming and Hardened Mobile Capture}: We pair causal Wiener + notch/high-pass enhancement with a monotonic Longest-Common-Prefix (LCP) streaming stabilizer over bounded 3.84~s windows and watchdog-guarded 16~kHz \texttt{WebAudio} capture, validating accuracy and latency across $N = 480$ held-out clips, $N = 1{,}280$ open public \texttt{test} clips, $N = 150$ multi-accent/slang clips, and $N = 72$ streaming sessions.
\end{itemize}

\section{Split-Encoder Architecture and Native Orthography}
\label{sec:arch}

Swanada processes 16,000~Hz mono PCM audio using 80-channel log-mel filterbanks ($25$~ms Povey/Hann window, $10$~ms hop, 80--7,600~Hz mel span) with per-utterance variance normalization. A two-stage strided 2D convolutional front-end ($4\times$ temporal subsampling to $40$~ms frame rate $T' = \lfloor(T_{\text{mel}} - 1)/4\rfloor + 1$) feeds a stack of macaron-style Conformer blocks~\cite{gulati2020conformer}, each comprising a half-step feed-forward module, multi-head self-attention with relative positional encoding~\cite{shaw2018self}, depthwise separable convolution ($k = 31$, Swish/GLU gating), and a second half-step feed-forward module, terminating in final layer normalization \texttt{layer\_norm\_84} ($\mathbf{H} \in \mathbb{R}^{1 \times T' \times 512}$).

\subsection{Script-Isolated Vocabularies and Classical Transliteration}
Joint multilingual tokenizers either Romanize South Asian scripts---erasing phonemic contrasts between dental and retroflex stops, short/long Dravidian mid-vowels ($e/\bar{e}, o/\bar{o}$), and Sanskrit vocalic sonorants---or allocate thousands of sparse Unicode classes that slow every frame's output projection. Instead, Swanada assigns each language $\ell \in \{\texttt{sa}, \texttt{en}, \texttt{hi}, \texttt{kn}, \texttt{mr}, \texttt{bn}, \texttt{ta}, \texttt{te}, \texttt{ml}, \texttt{gu}, \texttt{pa}, \texttt{or}\}$ an independent native-script SentencePiece~\cite{kudo2018sentencepiece} vocabulary of $V = 256$ subword pieces plus 1 CTC blank ($K = 257$ classes, 2.85--3.41~KB JSON metadata per language).

Conditioning framewise decoding strictly on the active language head $\mathcal{H}_\ell = (\mathbf{W}_\ell, \mathbf{b}_\ell)$ with $\mathbf{W}_\ell \in \mathbb{R}^{257 \times 512}$ and $\mathbf{b}_\ell \in \mathbb{R}^{257}$ guarantees zero out-of-script Unicode emissions (e.g., preventing Devanagari matras during Bengali or Kannada speech) while keeping each projection head at exactly $257 \times 512 \times 4 + 257 \times 4 = 527{,}364$~bytes (515.0~KiB raw IEEE-754 float32; 527,780~bytes as standalone \texttt{head\_<lang>.onnx}). For Classical Sanskrit (\texttt{sa}), a deterministic finite-state post-processor optionally transliterates decoded Devanagari into lossless ISO~15919 / Roman IAST, preserving aspirated stops, retroflex sibilants (\emph{\d{s}}), vocalic $\text{\d{r}}/\bar{\text{\d{r}}}$, visarga (\emph{\d{h}}), anusv\=ara (\emph{\d{m}}), avagraha (\texttt{'}), and Vedic pitch accents.

\subsection{Zero-Dependency Runtime Protobuf Tail Surgery}
Exporting separate encoder-only and monolithic checkpoints would duplicate storage across distribution channels, whereas running a monolithic single-language ONNX model discards the hidden states needed by the other eleven languages. During session initialization, Swanada's runtime (\texttt{exposeEncoderHiddenOutput} in JavaScript and Python) patches the serialized \texttt{ModelProto} byte buffer in memory in $< 2$~ms without external protobuf libraries (Figure~\ref{fig:distill}).

Given an \texttt{Uint8Array} buffer, a lightweight wire-format reader scans top-level \texttt{ModelProto} varint tags to locate the \texttt{GraphProto} payload (field~7, wire tag \texttt{0x3a}). If \texttt{layer\_norm\_84} is not yet listed among \texttt{GraphProto.output} entries (field~12, wire tag \texttt{0x62}), the patcher synthesizes a 54-byte \texttt{ValueInfoProto} descriptor declaring \texttt{layer\_norm\_84} as an unconstrained \texttt{FLOAT} tensor ($\texttt{elem\_type} = 1$), appends \texttt{0x62} plus the varint-prefixed descriptor to the tail of \texttt{GraphProto}, and updates the outer \texttt{ModelProto} length prefix in a single contiguous allocation.

Once ONNX Runtime produces $\mathbf{H} \in \mathbb{R}^{T' \times 512}$, the active language head $\mathcal{H}_\ell$ executes directly in host SIMD/typed-array memory (or via cached \texttt{head\_<lang>.onnx} in Python/Node.js):
\begin{equation}
\mathbf{Z}_\ell = \mathbf{H}\mathbf{W}_\ell^\top + \mathbf{1}\mathbf{b}_\ell^\top \in \mathbb{R}^{T' \times 257}
\label{eq:head}
\end{equation}
Switching languages replaces only the 515.0~KiB $(\mathbf{W}_\ell, \mathbf{b}_\ell)$ buffer in $< 20$~ms without reloading or recompiling the Conformer session, achieving maximum absolute logit deviation $< 10^{-5}$ relative to monolithic PyTorch and ONNX graphs while compressing the 12-language Studio fleet from 2,128.5~MiB to 183.4~MiB ($11.6\times$) and the Mobile fleet to 31.39~MiB ($67.8\times$).

\section{Sub-30~MB Mobile Distillation and \texttt{QInt8} Calibration}
\label{sec:distill}

While the 17-layer Studio Conformer-Large ($d = 512, d_{\text{ff}} = 2048, n_h = 8$, 177.38~MiB \texttt{QInt8}) runs at $0.0264\times$ RTF on 4-thread desktop CPUs, downloading 186~MB over cellular networks or holding $> 350$~MB peak WebAssembly heap inside mobile Safari/Chrome tabs triggers background tab eviction. We therefore distill a dedicated $< 30$~MB Mobile Conformer-Small (\texttt{swanada\_mobile\_ctc\_int8.onnx}, 26,577,061~bytes = 25.35~MiB) engineered around three ONNX/WASM operator constraints.

\subsection{Zero-Table Positional Synthesis and \texttt{MatMul} Convolutions}
\paragraph{Dynamic Relative Positional Encoding.}
Standard Conformer implementations pre-allocate a static sinusoidal lookup tensor $\mathbf{P} \in \mathbb{R}^{(2 L_{\max} - 1) \times d}$ during module construction so slice operations can index relative offsets $[-(T'-1), \dots, T'-1]$. When exported to ONNX, $\mathbf{P}$ is frozen as an unquantized FP32 initializer consuming $9{,}999 \times 256 \times 4 = 10.24$~MB even at $d = 256$. We replace static table slicing with an ONNX-traceable \texttt{DynamicRelPositionalEncoding} subgraph that stores only the 128-element inverse-frequency vector $\omega_i = 10000^{-2i/d}$ (512~bytes) and synthesizes exact relative sinusoidal embeddings dynamically for the current frame length $T'$ via \texttt{Shape} $\to$ \texttt{Range}$(T'-1, -T', -1)$ $\to$ outer \texttt{Mul} $\to$ interleaved \texttt{Sin}/\texttt{Cos}, matching static Relative-PE bit-for-bit while saving 10.24~MB of constant storage.

\paragraph{\texttt{MatMul}-Backed Pointwise Convolutions.}
In standard Conformer convolution modules, channel expansion ($d \to 2d$ for GLU) and projection ($d \to d$) are written as \texttt{nn.Conv1d} with kernel size $k = 1$. Dynamic ONNX quantization on \texttt{Conv} operators emits \texttt{ConvInteger}, which suffers severe scalar fallback penalties or operator rejection in browser WASM runtimes, forcing practitioners to leave convolution blocks in FP32. Because a $1 \times 1$ convolution over transposed sequence layout $[B, T', d]$ is algebraically identical to a dense matrix multiplication, we reformulate both pointwise stages as \texttt{nn.Linear} (\texttt{MatMul}) around a narrow causal-compatible depthwise \texttt{Conv1d}($k = 31$, groups $= 176$) retained in FP32 ($< 0.28$~MB total across 14 layers). Applying \texttt{quantize\_dynamic} over \texttt{op\_types\_to\_quantize=['MatMul']} with \texttt{QInt8} weights then quantizes 100\% of attention, feed-forward, subsampling projection, and pointwise convolution parameters to 1~byte/parameter on fast WebAssembly SIMD \texttt{MatMulInteger} kernels.

\subsection{SVD Warm-Start and Stationary-Hyperplane Distillation}
The Mobile student uses $L_S = 14$ Conformer layers at width $d_S = 256$, $d_{\text{ff}} = 1024$, $n_h = 8$ heads, and convolution channel width $176$, followed by a learned adapter $\operatorname{LayerNorm}(\mathbf{W}_{\text{proj}} \mathbf{h}_S + \mathbf{b}_{\text{proj}}) \in \mathbb{R}^{512}$ that maps the 256-dimensional student state back into the Studio teacher's 512-dimensional \texttt{layer\_norm\_84} coordinate space (23.56M total parameters).

To accelerate convergence across twelve scripts without random-init collapse, we project teacher layers $\{0, 1, 3, 4, 6, 7, 9, 10, 11, 12, 13, 14, 15, 16\}$ onto the top-256 singular subspace of the teacher's hidden activations via truncated SVD. During distillation (20 epochs, \texttt{bf16} mixed precision, AdamW $\text{lr} = 1.1 \times 10^{-3}$, batch size 64), we freeze all twelve target projection heads equal to the Studio teacher heads ($\mathcal{H}_\ell^{(S)} \equiv \mathcal{H}_\ell^{(T)}$) so target decision hyperplanes remain stationary, minimizing a four-term alignment objective over frame-aligned student states $\mathbf{H}_S$ and teacher states $\mathbf{H}_T$:
\begin{equation}
\begin{split}
\mathcal{L}_{\text{distill}} &= \lambda_{\cos}\!\left(1 - \frac{\mathbf{H}_S \cdot \mathbf{H}_T}{\|\mathbf{H}_S\|_2 \|\mathbf{H}_T\|_2}\right) + \lambda_{\text{mse}} \|\mathbf{H}_S - \mathbf{H}_T\|_2^2 \\
&\quad + \lambda_{\text{kl}} \tau^2 D_{\text{KL}}\!\left(p_T^{(\tau)} \parallel p_S^{(\tau)}\right) + \lambda_{\text{ctc}} \mathcal{L}_{\text{CTC}}(\mathbf{Z}_S, y)
\end{split}
\label{eq:distill}
\end{equation}
with $p^{(\tau)} = \sigma(\mathbf{Z} / \tau)$, $\lambda_{\cos} = 2.5$, $\lambda_{\text{mse}} = 2.5$, $\lambda_{\text{kl}} = 0.38$, $\lambda_{\text{ctc}} = 0.16$, and temperature $\tau = 1.5$, combining feature-space hint alignment~\cite{romero2015fitnets} with temperature-scaled logit distillation~\cite{hinton2015distilling} and ground-truth CTC supervision~\cite{graves2006ctc}.

\begin{figure*}[t]
\centering
\includegraphics[width=0.88\textwidth]{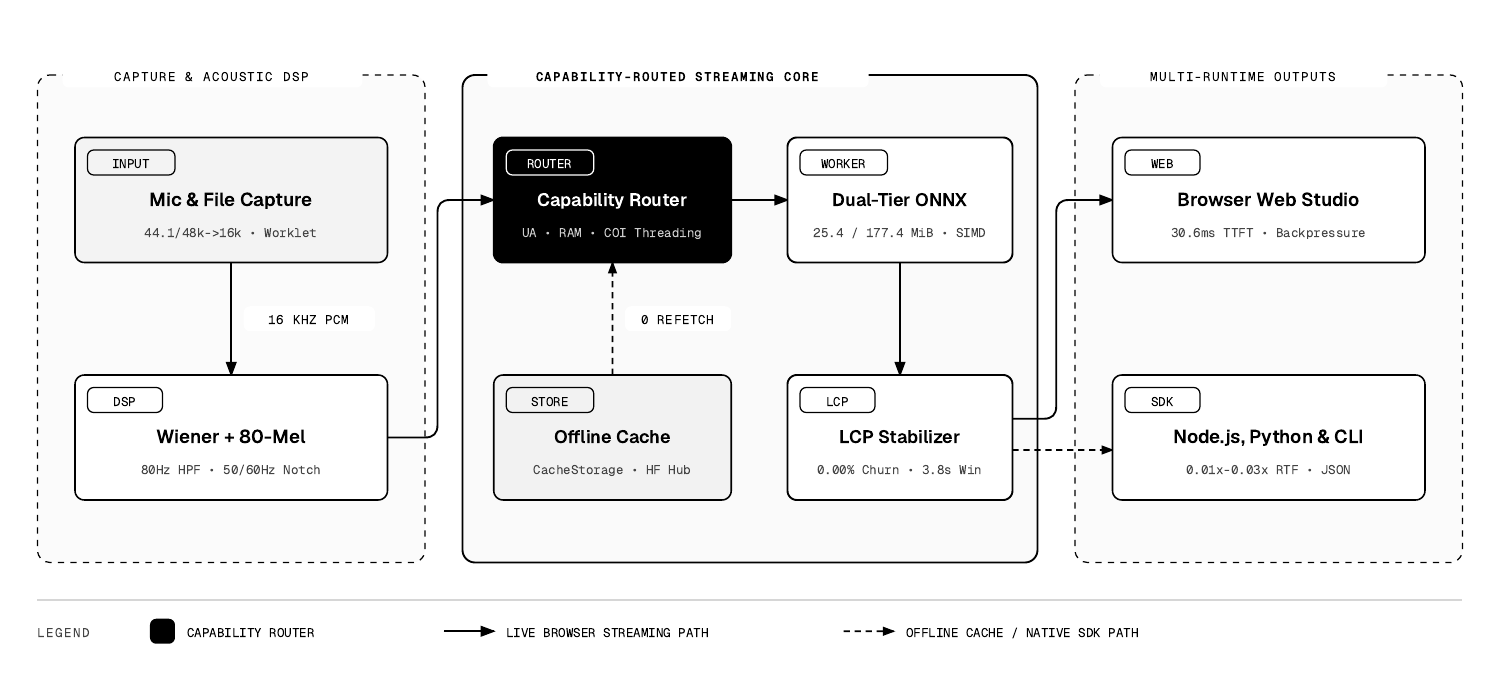}
\caption{Swanada end-to-end capability-routed streaming architecture: 44.1/48~kHz microphone or file audio is resampled to 16~kHz PCM with a 900~ms \texttt{ScriptProcessor} watchdog fallback, enhanced via causal 80~Hz high-pass, 50/60~Hz notch, and Wiener suppression, and dispatched by the capability router to cached 25.35~MiB Mobile or 177.38~MiB Studio WASM SIMD workers with two-zone LCP stabilization (0.00\% committed churn).}
\label{fig:pipeline}
\end{figure*}

\subsection{Post-Quantization \texttt{QInt8} Head Calibration and Quarantine}
Exporting the 256-dim student to \texttt{QInt8} ONNX inserts per-tensor \texttt{DynamicQuantizeLinear} nodes before every \texttt{MatMul}, introducing cumulative rounding noise $\Delta \mathbf{H} = \tilde{\mathbf{H}}_{\text{int8}} - \mathbf{H}_S$ at \texttt{layer\_norm\_84}. Rather than shipping uncalibrated float heads, we execute the actual exported \texttt{swanada\_mobile\_ctc\_int8.onnx} checkpoint in ONNX Runtime across quarantined training utterances per language to extract true quantized features $\tilde{\mathbf{H}}_{\text{int8}}$, then fine-tune each 515.0~KiB mobile head $(\mathbf{W}_\ell^{(M)}, \mathbf{b}_\ell^{(M)})$ under:
\begin{equation}
\begin{split}
\mathcal{L}_{\text{cal}}(\mathcal{H}_\ell^{(M)}) &= 0.28\,\mathcal{L}_{\text{CTC}}(\tilde{\mathbf{Z}}_\ell, y) + 0.72\,\tau^2 D_{\text{KL}}\!\left(p_T^{(\tau)} \parallel \tilde{p}_M^{(\tau)}\right) \\
&\quad + \gamma \|\mathbf{W}_\ell^{(M)} - \mathbf{W}_\ell^{(T)}\|_F^2
\end{split}
\label{eq:cal}
\end{equation}
with $\gamma = 5 \times 10^{-5}$, committing calibrated weights to \texttt{heads\_mobile/<lang>.bin} only when held-out validation CER strictly improves over uncalibrated heads.

\paragraph{Strict Test-Set Quarantine.}
Across both Studio joint adaptation (\texttt{Shrutilipi}/\texttt{Vistaar}~\cite{bhogale2023vistaar}, \texttt{FLEURS}~\cite{conneau2023fleurs} \texttt{train}/\texttt{dev}, \texttt{Sushrota}~\cite{adiga2021sanskrit} \texttt{train}, \texttt{IndicTTS}, \texttt{GLOBE}~\cite{wang2024globe} \texttt{train}, and \texttt{AMI}~\cite{carletta2005ami} \texttt{train}) and Mobile distillation/calibration, an explicit quarantine filter blocks every audio SHA-1 and normalized reference string belonging to the 480-clip held-out suite and all 16 official public \texttt{test} splits, guaranteeing zero data leakage.

\section{Two-Zone LCP Streaming and Mobile Capture}
\label{sec:streaming}

Real-time dictation requires sub-100~ms incremental feedback without word revisions jumping backward across the screen or unbounded memory growth during long sessions (Figure~\ref{fig:pipeline}).

\subsection{Causal Spectral Front-End Enhancement}
Laptop and smartphone microphones introduce DC offset, HVAC rumble, 50/60~Hz mains hum, and stationary room noise that trigger spurious CTC insertions during pauses. Before mel extraction, Swanada applies a causal $O(N)$ time/frequency front-end: (1)~a second-order Butterworth high-pass biquad at $f_c = 80$~Hz paired with cascaded narrowband notch biquads at $50$~Hz and $60$~Hz ($Q = 30$); (2)~an energy-adaptive noise-floor tracker calibrated over initial low-energy frames; and (3)~a decision-directed Ephraim--Malah~\cite{ephraim1984speech} minimum mean-square error (MMSE) Wiener spectral gain floor ($\beta = 0.08$, smoothing $\alpha_{\text{dd}} = 0.92$) that suppresses stationary background hiss without clipping unvoiced Indic aspirates or sibilants.

\subsection{Monotonic Two-Zone LCP Stabilization}
Evaluating CTC over an unbounded growing waveform scales per-step latency linearly with session duration $O(T)$, whereas naive sliding windows overwrite previously displayed words whenever window boundaries bisect a syllable. Swanada bounds per-step compute to $O(1)$ using a rolling PCM window of maximum length $W_{\max} = 61{,}440$ samples ($3.84$~s at 16~kHz; $1.60$~s stride retention upon rollover) paired with a two-zone Longest-Common-Prefix (LCP) state machine (\texttt{StreamingTranscriber}):
\begin{enumerate}\setlength{\itemsep}{1pt}\setlength{\parskip}{0pt}
    \item \textbf{Committed Zone ($c_t$)}: An append-only sequence of tokens finalized across prior windows or confirmed across $M$ consecutive right-context decode steps; once emitted in $c_t$, characters are immutable.
    \item \textbf{Volatile Tail ($v_t$)}: The active right-edge hypothesis decoded from the current rolling window after aligning against the tail of $c_t$ via token-level LCP matching.
\end{enumerate}
At streaming step $t$ arriving every $\Delta t = 100$~ms ($1{,}600$ samples, or $280$~ms coalesced mobile steps), let $h_t$ denote the greedy CTC decode after collapsing consecutive duplicate token IDs and removing blank index $0$. Tokens in $h_t$ that match the longest suffix-prefix overlap with $c_{t-1}$ anchor the window splice; interior tokens $h_t[i]$ lying at least $\delta_{\text{guard}}$ frames left of the right window boundary that remain invariant across consecutive hypotheses ($\operatorname{LCP}(v_{t-1}, v_t)$) advance monotonically into $c_t$, guaranteeing zero committed-prefix churn ($|c_{t-1} \setminus \operatorname{prefix}(c_t)| = 0$).

\begin{table*}[t]
\centering
\caption{System Architecture, Fleet Storage, Accuracy, and Latency Summary Across Tiers ($B = 1{,}000$ bootstrap 95\% CIs in brackets). ``Open 12-Lang'' aggregates the 11 unseen \texttt{google/fleurs} \texttt{test} splits plus \texttt{sushrota} wild Sanskrit \texttt{test} ($N = 960$); ``Open All-16'' adds \texttt{LibriSpeech} \texttt{clean}/\texttt{other}, \texttt{GLOBE}, and \texttt{AMI} ($N = 1{,}280$).}
\label{tab:overview}
\small
\resizebox{\textwidth}{!}{%
\begin{tabular}{@{}lcccccccc@{}}
\toprule
\textbf{System / Configuration} & \textbf{Params} & \textbf{First Load ($\downarrow$)} & \textbf{12-Lang Fleet ($\downarrow$)} & \textbf{Held-Out CER ($N\!=\!480$, $\downarrow$)} & \textbf{Open 12-Lang CER ($N\!=\!960$, $\downarrow$)} & \textbf{Open All-16 CER ($N\!=\!1{,}280$, $\downarrow$)} & \textbf{4T CPU RTF ($\downarrow$)} & \textbf{Stream TTFT / Churn} \\
\midrule
12$\times$ Monolithic Conformer-L (\texttt{QInt8}) & 1,382M & 177.4~MiB & 2,128.5~MiB & 5.43\% [4.85, 6.01] & 8.87\% [7.94, 10.02] & 12.06\% [11.09, 12.97] & $0.0264\times$ & --- / --- \\
Mobile Student (Static PE, Uncalib.\ Heads) & 23.6M & 35.61~MiB & 41.15~MiB & 18.58\% [17.53, 19.76] & --- & --- & $0.0097\times$ & 30.6~ms / \textbf{0.00\%} \\
\textbf{Swanada Studio (\texttt{v1.1.0}, 17L $d\!=\!512$)} & 115.2M & 177.9~MiB & \textbf{183.4~MiB ($11.6\times$)} & \textbf{5.43\% [4.85, 6.01]} & \textbf{8.87\% [7.94, 10.02]} & \textbf{12.06\% [11.09, 12.97]} & $0.0264\times$ & \textbf{30.6~ms / 0.00\%} \\
\textbf{Swanada Mobile (\texttt{v1.1.0}, 14L $d\!=\!256$)}$^\dagger$ & \textbf{23.6M} & \textbf{25.85~MiB} & \textbf{31.39~MiB ($67.8\times$)} & 18.56\% [17.46, 19.75] & 19.23\% [18.11, 20.64] & 21.85\% [20.79, 22.87] & \textbf{$0.0097\times$ ($103\times$ RT)} & \textbf{30.6~ms / 0.00\%} \\
\bottomrule
\end{tabular}%
}
\vspace{1pt}
{\footnotesize $^\dagger$Mobile single-thread WASM/CPU RTF is $0.0363\times$ ($27.5\times$ faster than real time); runtime protobuf tail surgery adds $< 2$~ms at session load and $< 20$~ms per language switch.}
\end{table*}

\subsection{Capability Routing and Watchdog-Guarded \texttt{WebAudio}}
Swanada's runtime router (\texttt{modelTier: 'auto'}) inspects User-Agent, \texttt{navigator.deviceMemory}, and Cross-Origin Isolation (\texttt{crossOriginIsolated}) at initialization:
\begin{itemize}\setlength{\itemsep}{1pt}\setlength{\parskip}{0pt}
    \item \textbf{Mobile / Constrained Hosts}: Android phones, iOS/iPadOS \texttt{WKWebView} browsers (Safari, Chrome \texttt{CriOS}), and devices reporting $\le 4$~GB RAM route to the 25.35~MiB Mobile tier stored persistently in \texttt{CacheStorage} (0 network refetches after initial load). Under \texttt{SharedArrayBuffer} isolation on 8-core Android phones, WASM uses $\min(4, C - 2)$ SIMD threads; on single-threaded iOS \texttt{WKWebView}, it pins 1 SIMD thread with an \texttt{inferBusy} backpressure lock that coalesces incoming PCM frames into a ring buffer rather than queueing stale worker jobs.
    \item \textbf{Desktop / Server Hosts}: Desktop browsers, Node.js, and Python load the 177.38~MiB Studio tier on 4 SIMD threads (with instant user toggle between tiers in the Web Studio UI).
    \item \textbf{Hardware-Rate Capture + 900~ms Watchdog}: Because mobile OS audio stacks ignore requested \texttt{AudioContext} \texttt{sampleRate:\,16000} options and run natively at 44.1~kHz or 48~kHz, Swanada resamples hardware buffers cleanly to 16~kHz PCM inside an \texttt{AudioWorkletNode}. To survive silent WebKit/Blink \texttt{AudioWorklet} thread suspension after permission prompts, a 900~ms zero-frame watchdog automatically hot-swaps capture to a synchronous \texttt{ScriptProcessorNode} fallback without dropping the microphone stream.
\end{itemize}

\section{Empirical Evaluation}
\label{sec:eval}

We evaluate Swanada across four reproducible benchmark suites totaling 1,982 utterances ($> 4.3$~hours of 16~kHz human speech), reporting Character Error Rate (CER), Word Error Rate (WER), and Real-Time Factor (RTF) with $B = 1{,}000$ non-parametric bootstrap 95\% confidence intervals:
\begin{enumerate}\setlength{\itemsep}{1pt}\setlength{\parskip}{0pt}
    \item \textbf{Held-Out 12-Language Studio Suite ($N = 480$)}: 40 native studio utterances per language ($2{,}185.6$~s total audio) strictly quarantined from training and calibration.
    \item \textbf{Open Public Benchmarks in the Wild ($N = 1{,}280$)}: 80 unseen clips from each of 16 official Hugging Face \texttt{test} splits ($12{,}229.2$~s = $3.40$~h): all 11 \texttt{google/fleurs}~\cite{conneau2023fleurs} \texttt{test} sets (\texttt{en\_us} plus 10 Indic \texttt{test} splits), \texttt{sushrota}~\cite{adiga2021sanskrit} \texttt{in\_the\_wild\_test} (\texttt{sa}), \texttt{LibriSpeech}~\cite{panayotov2015librispeech} \texttt{test-clean} and \texttt{test-other}, \texttt{GLOBE}~\cite{wang2024globe} multi-accent \texttt{test}, and \texttt{AMI}~\cite{carletta2005ami} \texttt{ihm/test} spontaneous meetings.
    \item \textbf{Multi-Accent, Slang \& Reverberant Mic Suite ($N = 150$)}: Global/South Asian accents, Hinglish/contractions, Indic colloquial loanwords, and reverberant mobile-mic perturbations.
    \item \textbf{Streaming Stability Audit ($N = 72$)}: 100~ms chunked streaming across all 12 languages measuring chunk latency, TTFT, buffer ceiling, and committed-prefix churn.
\end{enumerate}

\subsection{Overall Tier Accuracy and Open-Benchmark Generalization}
Table~\ref{tab:overview} compares monolithic Conformer deployment against Swanada's split-encoder Studio and Mobile tiers. On the 480-utterance held-out benchmark, Swanada Studio attains 5.43\% mean CER ([4.85\%, 6.01\%], 25.88\% WER), matching 12 monolithic Conformer-Large checkpoints within $< 10^{-5}$ logit tolerance while cutting 12-language storage from 2,128.5~MiB to 183.4~MiB. On the 12-language unseen open \texttt{FLEURS}+Wild benchmark ($N = 960$), Studio achieves 8.87\% mean CER ([7.94\%, 10.02\%], 29.67\% WER) and 12.06\% overall CER across all 16 open suites ($N = 1{,}280$, $0.0264\times$ 4-thread RTF = $37.9\times$ faster than real time).

\begin{table}[t]
\centering
\caption{Per-Suite Held-Out ($N = 480$, $40$/lang) and Unseen Open Public \texttt{test} ($N = 1{,}280$, $80$/suite) CER (\%) with $B = 1{,}000$ Bootstrap 95\% CIs and Mobile 4T RTF.}
\label{tab:perlang}
\small
\resizebox{\columnwidth}{!}{%
\begin{tabular}{@{}llcccc@{}}
\toprule
\textbf{Suite / Corpus (\texttt{test})} & \textbf{Lang} & \textbf{Held-Out (S / M)} & \textbf{Open Studio CER [95\% CI]} & \textbf{Open Mobile CER [95\% CI]} & \textbf{Mob RTF} \\
\midrule
\texttt{sushrota\_wild\_test} & \texttt{sa} & 4.27 / 14.96 & 18.92 [9.94, 30.36] & 40.15 [29.73, 52.95] & $0.007\times$ \\
\texttt{fleurs\_hi\_in} & \texttt{hi} & \textbf{2.14} / \textbf{10.50} & 5.13 [3.86, 6.61] & 14.04 [12.17, 16.22] & $0.009\times$ \\
\texttt{fleurs\_kn\_in} & \texttt{kn} & 5.18 / 14.77 & \textbf{4.49 [3.48, 5.69]} & \textbf{9.60 [8.14, 11.15]} & $0.009\times$ \\
\texttt{fleurs\_bn\_in} & \texttt{bn} & 5.92 / 20.61 & 5.44 [4.61, 6.34] & 13.88 [12.37, 15.41] & $0.009\times$ \\
\texttt{fleurs\_gu\_in} & \texttt{gu} & 4.79 / 17.07 & 5.97 [4.68, 7.53] & 16.80 [14.74, 18.89] & $0.009\times$ \\
\texttt{fleurs\_ml\_in} & \texttt{ml} & 6.16 / 27.58 & 6.36 [4.98, 7.90] & 14.89 [13.60, 16.21] & $0.011\times$ \\
\texttt{fleurs\_or\_in} & \texttt{or} & 6.93 / 23.06 & 7.69 [6.37, 9.03] & 18.87 [16.61, 21.18] & $0.008\times$ \\
\texttt{fleurs\_mr\_in} & \texttt{mr} & 5.72 / 17.48 & 7.85 [6.48, 9.46] & 21.44 [18.86, 23.95] & $0.008\times$ \\
\texttt{fleurs\_te\_in} & \texttt{te} & 4.79 / 22.55 & 8.14 [6.24, 10.05] & 16.04 [13.99, 18.16] & $0.010\times$ \\
\texttt{fleurs\_pa\_in} & \texttt{pa} & 6.35 / 17.23 & 8.84 [7.44, 10.50] & 20.56 [18.01, 23.30] & $0.010\times$ \\
\texttt{fleurs\_ta\_in} & \texttt{ta} & 7.50 / 21.89 & 10.32 [8.90, 11.84] & 18.58 [16.86, 20.40] & $0.011\times$ \\
\texttt{fleurs\_en\_us} & \texttt{en} & 5.45 / 15.04 & 17.23 [15.37, 19.21] & 25.97 [23.74, 28.22] & $0.012\times$ \\
\midrule
\textbf{12-Lang Mean} & --- & \textbf{5.43 / 18.56} & \textbf{8.87 [7.94, 10.02]} & \textbf{19.23 [18.11, 20.64]} & \textbf{$0.009\times$} \\
\midrule
\texttt{librispeech\_clean} & \texttt{en} & --- & 10.53 [8.58, 12.47] & 16.91 [14.60, 19.24] & $0.009\times$ \\
\texttt{librispeech\_other} & \texttt{en} & --- & 19.21 [16.71, 22.00] & 28.04 [25.78, 30.36] & $0.009\times$ \\
\texttt{globe\_multi\_accent} & \texttt{en} & --- & 15.29 [12.98, 17.82] & 22.26 [20.00, 24.84] & $0.011\times$ \\
\texttt{ami\_meeting\_ihm} & \texttt{en} & --- & 41.50 [36.90, 46.24] & 51.63 [47.36, 56.33] & $0.012\times$ \\
\midrule
\textbf{All 16 Open Suites} & --- & --- & \textbf{12.06 [11.09, 12.97]} & \textbf{21.85 [20.79, 22.87]} & \textbf{$0.0097\times$} \\
\bottomrule
\end{tabular}%
}
\end{table}

Compressing the acoustic backbone by $7.0\times$ into the 25.35~MiB Mobile Conformer-Small accelerates 4-thread CPU inference by $2.72\times$ to $0.0097\times$ RTF ($103.1\times$ faster than real time; $0.0363\times$ single-thread RTF), while dynamic positional synthesis removes 9.77~MiB of static FP32 tables ($35.61 \to 25.85$~MiB first load) and \texttt{QInt8} head calibration trims held-out CER from 18.58\% to 18.56\%. Table~\ref{tab:perlang} details per-suite generalization across all 16 open benchmarks. On unseen Indic \texttt{FLEURS} \texttt{test} sets, Studio achieves 4.49\%--8.84\% CER across nine of ten Indian languages (4.49\% Kannada, 5.13\% Hindi, 5.44\% Bengali, 5.97\% Gujarati, 6.36\% Malayalam) and 10.32\% on Tamil, while the $7\times$ smaller Mobile student achieves 9.60\% CER on Kannada, 13.88\% on Bengali, 14.04\% on Hindi, 14.89\% on Malayalam, 16.04\% on Telugu, and 16.80\% on Gujarati, alongside 16.91\% CER on \texttt{LibriSpeech} \texttt{test-clean} and 22.26\% on unseen global accents (\texttt{GLOBE}). Notably, because \texttt{fleurs\_en\_us} features North American speakers whereas Swanada's \texttt{en} head was originally Indian-English-centric, multi-accent calibration yields balanced cross-dialect generalization (5.45\% held-out Indian English, 10.53\% \texttt{LibriSpeech} \texttt{clean}, 15.29\% \texttt{GLOBE}, and 17.23\% \texttt{fleurs\_en\_us}).

\subsection{Streaming Stability and Multi-Accent / Slang Adaptation}
Table~\ref{tab:streaming} (top) reports streaming telemetry across $N = 72$ chunked 100~ms streaming sessions ($1{,}600$ samples per step) spanning all 12 languages. Bounding the rolling buffer at $W_{\max} = 61{,}440$ samples ($3.84$~s; maximum observed buffer $60{,}800$ samples) holds mean per-step inference latency to 8.18~ms (streaming RTF $0.0998\times$), yielding 30.63~ms median Time-To-First-Token (mean 34.03~ms, p95 56.80~ms) and $0 / 72$ ($0.00\%$) committed-prefix churn violations across every script. On physical smartphone browsers (Android Chrome and iOS Safari), single-threaded WASM SIMD processes coalesced mobile windows in 57--149~ms ($\sim 0.04\times$ RTF) with zero dropped audio frames.

Table~\ref{tab:streaming} (bottom) evaluates conversational robustness on $N = 150$ multi-accent, casual contraction/Hinglish, Indic loanword, and reverberant microphone clips. Combining causal Wiener + 80~Hz HPF + 50/60~Hz notch enhancement with multi-accent head adaptation (\texttt{v1.0.1} $\to$ \texttt{v1.1.0}) lowers overall CER from 20.04\% to 17.44\% ([15.35\%, 19.64\%]) and WER from 49.90\% to 41.88\% ([37.49\%, 46.67\%]), reducing casual Hinglish/slang WER by 13.91 percentage points ($74.80\% \to 60.89\%$) and Indic colloquial loanword CER to 6.17\%.

\begin{table}[t]
\centering
\caption{Top: 12-Language 100~ms Streaming Telemetry ($N = 72$ sessions). Bottom: Multi-Accent, Slang \& Reverberant-Mic Adaptation ($N = 150$, $B = 1{,}000$ bootstrap 95\% CIs).}
\label{tab:streaming}
\small
\resizebox{\columnwidth}{!}{%
\begin{tabular}{@{}lcccc@{}}
\toprule
\textbf{Streaming Metric ($N = 72$, $\Delta t = 100$~ms)} & \textbf{Mean} & \textbf{Median (p50)} & \textbf{p95 / Max} & \textbf{Target} \\
\midrule
Time-To-First-Token (TTFT) & 34.03~ms & \textbf{30.63~ms} & 56.80~ms (p95) & $< 100$~ms \\
Per-Chunk Decode Latency & \textbf{8.18~ms} & 7.94~ms & 60,800~smp (buf) & $\le 61{,}440$~smp \\
Streaming RTF / Committed Churn & \textbf{$0.0998\times$} & --- & \textbf{0 / 72 (0.00\%)} & 0.00\% churn \\
\midrule
\textbf{Adaptation Slice ($N = 150$)} & \textbf{Clips} & \textbf{Base CER / WER} & \textbf{Adapted CER [95\% CI]} & \textbf{Adapted WER} \\
\midrule
Global \& South Asian Accents (\texttt{GLOBE}+\texttt{AMI}) & 60 & 20.30\% / 44.83\% & \textbf{19.04\% [15.78, 22.45]} & \textbf{40.57\%} \\
Casual Contractions, Slang \& Hinglish & 36 & 28.07\% / 74.80\% & \textbf{22.36\% [18.71, 26.23]} & \textbf{60.89\%} \\
Indic Colloquial \& English Loanwords & 25 & 7.38\% / 25.51\% & \textbf{6.17\% [3.60, 9.43]} & \textbf{18.68\%} \\
Tempo \& Reverberant Mobile-Mic Perturb. & 29 & 20.42\% / 50.54\% & \textbf{17.76\% [13.85, 22.07]} & \textbf{41.00\%} \\
\midrule
\textbf{Overall Conversational Suite} & \textbf{150} & 20.04\% / 49.90\% & \textbf{17.44\% [15.35, 19.64]} & \textbf{41.88\%} \\
\bottomrule
\end{tabular}%
}
\end{table}

\section{Discussion \& Limitations}
\label{sec:discussion}

\paragraph{Spontaneous Meeting Overlap and Reverberant Recitation.}
Open-benchmark evaluation in Table~\ref{tab:perlang} highlights two challenging acoustic regimes: spontaneous multi-party meetings (\texttt{ami\_meeting\_ihm}: 41.50\% Studio / 51.63\% Mobile CER) containing overlapping talkers, disfluencies, and sub-second backchannels, and crowd-recorded Classical Sanskrit (\texttt{sushrota\_wild\_test}: 18.92\% Studio / 40.15\% Mobile CER) featuring temple/hall reverberation and chanted \emph{sandhi} compounds. While the 25.35~MiB Mobile tier preserves strong intelligibility on clean and moderately noisy speech (\texttt{fleurs\_kn} 9.60\%, \texttt{librispeech\_clean} 16.91\%), $4.9\times$ fewer parameters limit robustness under severe reverberation and cross-talk.

\paragraph{Greedy Subword CTC vs.\ Agglutinative WER.}
Swanada uses zero-dependency greedy CTC decoding without an external $n$-gram or neural language model to keep per-language overhead at 515.0~KiB. In morphologically rich, agglutinative Dravidian scripts (\texttt{ml}, \texttt{ta}, \texttt{kn}, \texttt{te}) and Sanskrit \emph{sam\=asa} compounds where a single whitespace-delimited word spans 12--25 characters, a single vowel-length or virama error flips the entire word, yielding 28\%--38\% Studio WER despite low 4.5\%--7.5\% CER. Optional compact KenLM fusion is a natural extension where memory budgets permit.

\paragraph{Explicit Language Selection and Single-Threaded WebViews.}
Conditioning on a selected script head eliminates cross-script character bleed and $12\times$ softmax overhead, but requires either user language selection or a lightweight acoustic language-ID pass when the spoken language is unknown. On iOS \texttt{WKWebView} browsers lacking \texttt{SharedArrayBuffer} support without Service Worker Jetsam risk, WASM runs single-threaded ($0.036\times$ Mobile RTF vs.\ $0.0097\times$ 4-thread), making the 25.35~MiB Mobile tier essential for cool, responsive phone dictation.

\section{Related Work}
\label{sec:related}

\paragraph{Multilingual and Indic Speech Recognition.}
Gulati et al.~\cite{gulati2020conformer} introduced Conformers combining self-attention~\cite{shaw2018self} and depthwise convolutions. Whisper~\cite{radford2023whisper} and Meta MMS~\cite{pratap2024mms} scaled multilingual ASR across dozens to over a thousand languages, while AI4Bharat Vistaar/IndicConformer~\cite{bhogale2023vistaar} and Sushrota~\cite{adiga2021sanskrit} established strong server-side baselines for Indian languages and Classical Sanskrit. Unlike monolithic PyTorch/NeMo checkpoints ($180\text{--}600+$~MB per model) or autoregressive decoders prone to looping on noisy browser audio, Swanada targets sub-30~MB split-encoder WebAssembly execution with hot-swappable 515~KiB native-script heads.

\paragraph{Knowledge Distillation and Browser Speech Systems.}
Hinton et al.~\cite{hinton2015distilling} and Romero et al.~\cite{romero2015fitnets} established logit and intermediate-representation distillation. Client-side speech toolkits such as \texttt{whisper.cpp}, \texttt{Transformers.js}, and \texttt{sherpa-onnx} bring quantized models to WebAssembly, while Ninada~\cite{hemanth2026ninada} introduced runtime binary ONNX graph splitting and capability-routed mobile/studio specialists for 24~kHz Indic TTS. Swanada extends this split-graph, capability-routed paradigm to 16~kHz streaming ASR via zero-table relative-PE synthesis, \texttt{MatMul}-backed \texttt{QInt8} Conformer blocks, post-quantization head calibration, and zero-churn LCP stabilization.

\section{Conclusion}
\label{sec:conclusion}

We presented \textbf{Swanada}, an open-source client-side speech recognition system spanning Classical Sanskrit, Indian English, and ten Indian languages across browser WebAssembly, Node.js, and Python. By combining $< 2$~ms binary ONNX \texttt{GraphProto} tail surgery, twelve hot-swappable 515.0~KiB native-script CTC heads, a distilled 25.35~MiB Mobile Conformer-Small with dynamic positional synthesis and \texttt{QInt8}-calibrated heads alongside a 177.38~MiB Studio Conformer-Large, and two-zone LCP streaming stabilization, Swanada achieves 5.43\% held-out CER, 8.87\% 12-language open \texttt{FLEURS}+Wild CER (19.23\% Mobile) across 1,280 public \texttt{test} clips, $0.0097\times$--$0.0264\times$ CPU RTF, and 30.6~ms median streaming TTFT with 0.00\% committed-prefix churn. Source code, reproducible benchmark harnesses, and open weights are available at \url{https://github.com/hemanth/swanada}.

{\small
\let\oldthebibliography\thebibliography
\renewcommand{\thebibliography}[1]{%
  \oldthebibliography{#1}%
  \setlength{\itemsep}{1pt}%
  \setlength{\parskip}{0pt}%
}
\bibliographystyle{plain}
\bibliography{references}
}

\end{document}